\documentclass[aps,prl,twocolumn,showpacs,preprintnumbers,amsmath,amssymb,superscriptaddress]{revtex4-2}
\usepackage{graphicx}
\usepackage{xcolor}
\usepackage{hyperref}

\begin{document}

\title{Quantum-metric-driven light-induced ferrovalley state in d-wave altermagnets}

\author{Shihao Zhang}
\affiliation{School of Physics and Electronics, Hunan University, Changsha 410082, China}
\email{zhangshh@hnu.edu.cn}

\begin{abstract}
Isolating the quantum metric from the Berry curvature remains a central challenge in quantum materials, as the two geometric quantities nearly always coexist and their contributions are difficult to disentangle.
We show that d-wave altermagnets, whose real Hamiltonian possesses strictly vanishing Berry curvature, constitute an ideal platform for overcoming this obstacle.
Using the Magnus expansion and exact Floquet diagonalization, we demonstrate that linearly polarized off-resonant light drives an orbital-selective ferrovalley phase through a purely quantum-metric--mediated band-gap renormalization, with no Berry curvature contribution at any order.
The orbital selectivity originates from the hopping anisotropy, which generates a pronounced metric anisotropy between the $d_{xz}$ and $d_{yz}$ orbitals, and the gap reduction is expressed analytically in terms of the quantum metric.
The resulting valley gap difference provides a direct, quantitative measure of the quantum metric, accessible to spin-resolved ARPES and optical pump-probe spectroscopy.
This establishes d-wave altermagnets as a pristine, tunable platform in which quantum metric effects can be isolated, controlled by light polarization, and read out through valley polarization.


\end{abstract}
\maketitle

\textit{Introduction.} Quantum geometry, encompassing the quantum metric and Berry curvature, underpins a broad class of nonlinear optical and topological phenomena in condensed matter \cite{Ahn2022,Holder2020,Lindner2011}. 
The real part $g_{\mu\nu}$, the quantum metric, governs wavefunction overlaps in momentum space and has been linked to effects ranging from superfluid stiffness to nonlinear Hall responses.
The imaginary part $\Omega_{\mu\nu}$, the Berry curvature, drives anomalous transport and topological invariants.
While Berry curvature effects are now routinely detected and engineered, isolating \emph{pure} quantum metric signals has remained an outstanding challenge \cite{Gao2014,Sodemann2015,Ma2019}, because the metric and curvature nearly always coexist and their contributions interfere, making independent experimental signatures difficult to disentangle.

\begin{figure}[t]
\centering
\includegraphics[width=\columnwidth]{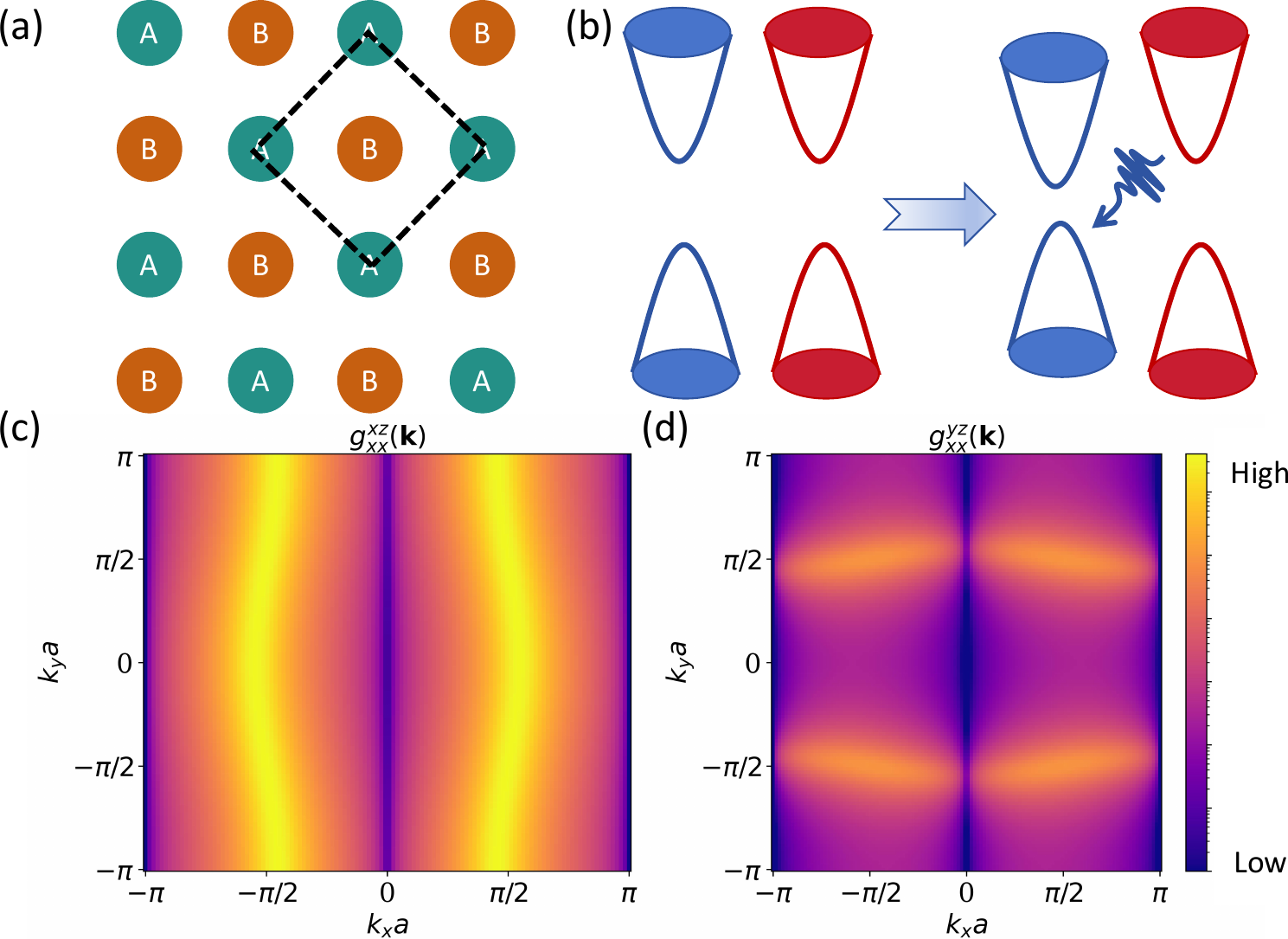}
\caption{Schematic of the quantum-metric-driven ferrovalley mechanism. (a) Square-lattice d-wave altermagnet with sublattices A (green) and B (red). (b) Under Floquet driving of $x$-direction linear polarized light, the $d_{xz}$ gap is remarkably reduced while the $d_{yz}$ gap remains essentially unchanged, producing a ferrovalley state with pronounced gap imbalance. Quantum metric $g_{xx}(\mathbf{k})$ in the first Brillouin zone for (c) $d_{xz}$ and (d) $d_{yz}$ orbitals with $V_\pi=1.0$\,eV, $V_\delta=0.15$\,eV, $|D_s|=0.25$\,eV.}
\label{fig:qmetric}
\end{figure}

Altermagnets\cite{Smejkal2022a,Smejkal2022b,aV2Se2O,chen2023spin,xiao2023spin,guo2024valley,zhang2025sliding,PhysRevLett.133.206401,fe2se2o,Zou_2025,li2025ferrovalley,c2fq-hkk4,tan2024bipolarized,duan2025antiferroelectric,PhysRevB.110.014442,PhysRevB.102.144441,zhang2025intrinsic,PhysRevB.110.064426,xu2025chemical,PhysRevB.110.144412,PhysRevLett.133.056401,zhang2026topological} have recently emerged as a uniquely suited platform for overcoming this obstacle. Their $d$-wave symmetric, real Hamiltonians enforce strictly vanishing Berry curvature while retaining a nontrivial quantum metric---a rare combination that naturally disentangles the two geometric contributions. The quantum metric of d-wave altermagnets has rapidly become an active area of investigation: recent theoretical works have studied quantum metric-induced nonlinear optical response\cite{zhang2026third,ezawa2026quantum} and nonreciprocal transport\cite{ouyang2025quantum} in altermagnets, quantum metric dipoles\cite{fang2023quantum} and their contribution to the nonlinear Hall effect, and the role of quantum geometry in altermagnetic tunnel junctions. 
Whether the quantum metric can drive observable electronic transitions under external driving fields in d-wave altermagnets---and yield unambiguous signatures without Berry curvature's influence---remains an open question.

In this work, we fill this gap by developing a Floquet theory of light-induced band-gap renormalization in d-wave altermagnets driven purely by the quantum metric---establishing a ferrovalley phase whose valley gap difference directly encodes the quantum metric anisotropy.
Starting from a minimal two-orbital tight-binding Hamiltonian on the square lattice, we derive the second-order quasi-energy correction using the Magnus expansion with the Peierls substitution, which reveals a direct, analytic proportionality between the driving field and the quantum metric, with no Berry curvature contribution at any order. Exact numerical Floquet diagonalization corroborates the analytical results and provides a non-perturbative benchmark, confirming that the predicted ferrovalley signal is robust beyond the perturbative regime.



\textit{Theoretical model and quantum geometry.} We consider a two-orbital ($d_{xz}$, $d_{yz}$) tight-binding model for d-wave altermagnets on a square lattice\cite{Vila2025}.
In the basis of sublattice ($\sigma$), orbital ($\tau$), and spin ($s$) degrees of freedom, the full static Hamiltonian is $\mathcal{H}_0 = \mathcal{H}_0^{\rm hop} + \mathcal{H}_{\rm CF} + \mathcal{H}_{\rm ex}$, where
\begin{align}
\mathcal{H}_0^{\rm hop} &= \sigma_x \otimes \left[A(k)\tau_0 + B(k)\tau_z\right] \otimes s_0, \\
\mathcal{H}_{\rm CF} &= \Delta\,\sigma_z \otimes \tau_z \otimes s_0, \\
\mathcal{H}_{\rm ex} &= m\,\sigma_z \otimes \tau_0 \otimes s_z,
\end{align}
with $A(k)=\frac{V_\pi+V_\delta}{2}(\cos k_x a+\cos k_y a)$ and $B(k)=\frac{V_\pi-V_\delta}{2}(\cos k_x a-\cos k_y a)$.
Here $V_\pi$ ($V_\delta$) are Slater-Koster $\pi$- ($\delta$-) bond integrals, $\Delta$ is the crystal-field splitting between $d_{xz}$ and $d_{yz}$, $m$ is the staggered exchange field, $s_z=\pm 1$ in the spin subspace, and $a$ is the sublattice spacing.
The $\sigma_x$ hopping term couples opposite sublattices; $\mathcal{H}_{\rm CF}$ splits the two orbitals on each sublattice; and $\mathcal{H}_{\rm ex}$ produces the spin-dependent staggered exchange field characteristic of altermagnetic order.

Since $[\mathcal{H}_0, s_z]=0$, the Hamiltonian block-diagonalizes into spin sectors.
Within each spin sector, the absence of direct $d_{xz}$--$d_{yz}$ hopping yields two independent $2\times2$ blocks (one per orbital), each acting on the sublattice degree of freedom:
\begin{equation}
\mathcal{H}_s^{\alpha}(k) = C^{\alpha}(k)\sigma_x + D_s^{\alpha}\sigma_z,
\end{equation}
where $\sigma_{x,z}$ are Pauli matrices acting on the two sublattice degrees of freedom (denoted A and B), and
\begin{align}
C^{xz}(k) &= V_\pi \cos k_x a + V_\delta \cos k_y a, \quad
D_s^{xz} = \frac{m s}{2} + \Delta, \\
C^{yz}(k) &= V_\delta \cos k_x a + V_\pi \cos k_y a, \quad
D_s^{yz} = \frac{m s}{2} - \Delta.
\end{align}
With $m=1.0$\,eV and $\Delta=0.25$\,eV, the minimal exchange splittings are $|D_s^{xz}|=|D_s^{yz}|=|m/2-\Delta|=0.25$\,eV, occurring for opposite spin species: spin-down ($s=-1$) for $d_{xz}$ and spin-up ($s=+1$) for $d_{yz}$. The high-gap branches ($|D_s|=0.75$\,eV) remain insulating at all relevant fields and are omitted hereafter; $D_s^{xz}$ and $D_s^{yz}$ refer exclusively to these low-gap branches. The zero-field gap is $E_g^0 = 2|D_s|=0.50$\,eV. The eigenvalues are $\pm h^{\alpha}(k)$ with $h^{\alpha}(k)=\sqrt{[C^{\alpha}(k)]^2+(D_s^{\alpha})^2}$. As shown in Fig.~2(a-b), the bandgaps of $d_{xz}$ and $d_{yz}$ bands are equal which exhibits the paravalley phase. 

The quantum geometric tensor $Q_{\mu
u}=g_{\mu
u}-\frac{i}{2}\Omega_{\mu
u}$ decomposes into the quantum metric $g_{\mu
u}$ (real, symmetric) and Berry curvature $\Omega_{\mu
u}$ (imaginary, antisymmetric). Since $\mathcal{H}_s^{\alpha}(k)$ is purely real, its eigenstates can be chosen real, forcing $\Omega_{\mu
u}(k)=0$ identically over the entire Brillouin zone. The quantum metric component is
\begin{equation}
g_{xx}^{\alpha}(k) = \frac{a^2 (D_s^{\alpha})^2 [\partial_{k_x}C^{\alpha}(k)]^2}{4[h^{\alpha}(k)]^4}.
\end{equation}
For the $d_{xz}$ orbital: $\partial_{k_x}C^{xz}=-a V_\pi\sin(k_x a)$, giving $g_{xx}^{xz} \propto V_\pi^2 \sin^2(k_x a)$. The $d_{yz}$ orbital has $g_{xx}^{yz} \propto V_\delta^2 \sin^2(k_x a)$. Since $V_\pi/V_\delta \sim 5$--$10$, the quantum metric is strongly orbital-anisotropic: $g_{xx}^{xz} \gg g_{xx}^{yz}$ as shown in Fig.~\ref{fig:qmetric}, which is the origin of the orbital-selective gap reduction derived below.

Having established the quantum geometry of the model, we now turn to how off-resonant light couples to these geometric degrees of freedom. 
The key insight is that the Peierls substitution $k_\mu \to k_\mu + (eE_\mu/\hbar\omega)\sin(\omega t)$ shifts the momentum in the $k_x$--$k_y$ plane, and the resulting $k$-space modulation of the Hamiltonian directly probes the wavefunction overlap between neighboring momentum states---precisely the information encoded in the quantum metric.
Because the Berry curvature vanishes identically, the leading interband coupling under periodic driving is exclusively geometric, governed by $g_{\mu\nu}$ rather than by the conventional Berry curvature dipole or anomalous velocity terms.

\textit{Floquet coupling to the quantum metric.} The Peierls substitution $k_\mu \to k_\mu + (eE_\mu/\hbar\omega)\sin(\omega t)$ couples the Floquet problem to $k$-space derivatives of the Hamiltonian. For a real two-band system $\mathcal{H}=h_x\sigma_x+h_z\sigma_z$, the interband matrix element of $\partial_{k_\mu}\mathcal{H}$ follows directly from the eigenstate structure. 
We can obtain the exact relation
\begin{equation}
|\langle u_-|\partial_{k_\mu}\mathcal{H}|u_+\rangle|^2 = 4h^2(k)\,g_{\mu\mu}(k),
\label{eq:interband_metric}
\end{equation}
which connects the interband coupling under momentum-space driving directly to the quantum metric.

Consequently, the quantum metric enters the Magnus expansion as the natural geometric quantity measuring interband coupling---a generic feature of real two-band Floquet systems with vanishing Berry curvature.

For $x$-polarized light $\mathbf{E}(t)=E_0\hat{x}\cos(\omega t)$, the Peierls substitution $k_x \to k_x + (eE_0/\hbar\omega)\sin(\omega t)$ yields the time-periodic Hamiltonian
\begin{equation}
\mathcal{H}(k,t) = C^{xz}\!\left(k_x + \frac{eE_0}{\hbar\omega}\sin(\omega t),\,k_y\right)\sigma_x + D_s\sigma_z,
\end{equation}
where $\alpha \equiv eE_0 a/\hbar\omega$ is the dimensionless driving parameter. $\mathcal{H}(k,t)$ remains real for all $t$, ensuring that the instantaneous and Floquet Berry curvature vanish identically.

We construct the effective Floquet Hamiltonian using the Magnus expansion~\cite{Magnus1954,Blanes2009,Martinez2003}, which expresses the time-ordered evolution operator over one period as $U(T,0)=\mathcal{T}\exp[-i\int_0^T \mathcal{H}(t)dt/\hbar] \equiv \exp[-i\mathcal{H}_F T/\hbar]$. The Magnus expansion converges for $\alpha \lesssim \pi$ and resums the Bessel-function dependence $J_n(\alpha)$ non-perturbatively through the Jacobi-Anger identity $\cos(k_x a + \alpha\sin\omega t) = \sum_n J_n(\alpha)\cos(k_x a + n\omega t)$.

The zeroth-order Magnus Hamiltonian describes dynamical band narrowing:
\begin{equation}
\mathcal{H}^{(0)}(k) = \bigl[V_\pi J_0(\alpha)\cos k_x a + V_\delta\cos k_y a\bigr]\sigma_x + D_s\sigma_z,
\end{equation}
where $J_0(\alpha) \leq 1$ reduces the effective $\pi$-bond hopping. At the valence band maximum or conduction band minimum, $h_x^{\rm eff}=0$, so the gap remains $2|D_s|$ at zeroth order. The first-order Magnus term $[\mathcal{H}^{(1)} \propto \sigma_y]$ also vanishes at the valence band maximum or conduction band minimum. The leading gap correction therefore arises at second order.

The second-order Magnus correction to the $\sigma_z$ component, which governs gap renormalization, is:
\begin{equation}
\mathcal{H}^{(2)}_{\sigma_z} = -\sum_{m\neq 0} \frac{2 D_s}{(m\hbar\omega)^2}\,
|\langle u_-|\mathcal{H}_m|u_+\rangle|^2\,\sigma_z,
\label{eq:magnus_main}
\end{equation}
where the interband matrix element $|\langle u_-|\mathcal{H}_m|u_+\rangle|^2$---\emph{not} the bare Fourier coefficient $|\mathcal{H}_m|^2$---appears because the double commutator $[[\mathcal{H}_{-m},\mathcal{H}_0],\mathcal{H}_m]$ in the Magnus series projects onto virtual interband transitions (see Appendix~\ref{app:magnus} for the explicit band-basis derivation).
Physically, the gap renormalization arises from an electron virtually absorbing a photon (via $\mathcal{H}_m$) to excite from the valence to the conduction band, then emitting a photon (via $\mathcal{H}_{-m}$) to return; the net energy shift is proportional to the interband transition amplitude squared.

\begin{figure}[t]
\centering
\includegraphics[width=\columnwidth]{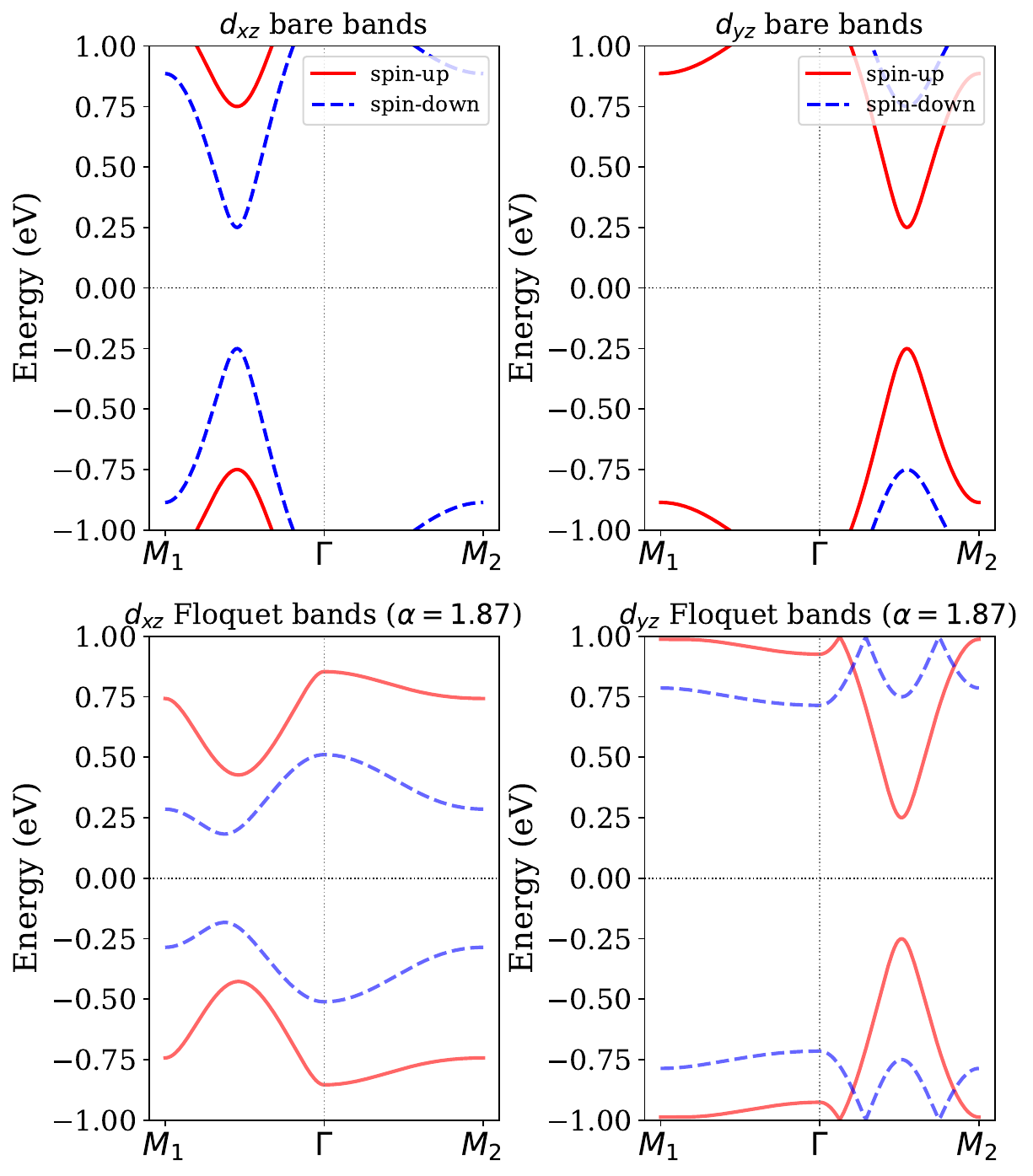}
\caption{Floquet driving electronic structures. The energy bands of (a) d$_{xz}$ orbit and (b) d$_{xz}$ orbit without Floquet driving, and Floquet bands with driving strength $\alpha=1.87$. Here $V_\pi=1.0$\,eV, $V_\delta=0.15$\,eV, $|D_s|=0.25$\,eV, and $\hbar \omega=2.0$\,eV.}
\label{fig:gap_evo}
\end{figure}

To connect this to the quantum metric, we exploit the exact proportionality between $\mathcal{H}_m$ and $\partial_{k_x}\mathcal{H}$ at the valence band maximum.
For odd $m$ (the only nonvanishing channels at the valence band maximum or conduction band minimum where $C=0$), the Jacobi--Anger expansion gives $\mathcal{H}_m^{\rm odd}=i\,\mathrm{sgn}(m)V_\pi J_{|m|}(\alpha)\sin(k_x a)\,\sigma_x$, while
$\partial_{k_x}\mathcal{H}=-a V_\pi\sin(k_x a)\,\sigma_x$.
Hence $\mathcal{H}_m^{\rm odd} = -i\,\mathrm{sgn}(m)(J_{|m|}(\alpha)/a)\,\partial_{k_x}\mathcal{H}$, an \emph{exact} proportionality (not a leading-order approximation) valid at all $k$.
The interband matrix element then factorizes:
\begin{equation}
|\langle u_-|\mathcal{H}_m|u_+\rangle|^2 = \frac{J_m^2(\alpha)}{a^2}\,
|\langle u_-|\partial_{k_x}\mathcal{H}|u_+\rangle|^2.
\end{equation}

Evaluating $|\langle u_-|\partial_{k_x}\mathcal{H}|u_+\rangle|^2$ from the eigenstates (see Eq.~\ref{eq:interband_metric} and Appendix~\ref{app:qmetric}) gives $4h^2(k)g_{xx}(k)$, leading to the explicit quantum-geometric form of the Magnus gap correction:
\begin{equation}
\Delta E_g^{\rm Magnus} = -\frac{16 D_s^3}{a^2 (\hbar\omega)^2}\,g_{xx}\,\sum_{m\neq 0} \frac{J_m^2(\alpha)}{m^2},
\label{eq:magnus_gap}
\end{equation}
evaluated at the valence band maximum or conduction band minimum. Equation~(\ref{eq:magnus_gap}) is the central analytical result of this work: the Floquet gap correction couples directly and exclusively to the quantum metric through the interband channel, with identically vanishing Berry curvature at all orders. In other words, the Peierls substitution implements a periodic shift of the electronic crystal momentum in k-space, with the k-space gradient operator $\partial_k$ acting as the corresponding infinitesimal generator. From this, one can rigorously deduce that the Fourier harmonics of the time-periodic Hamiltonian are proportional to the k-space derivatives of the static Hamiltonian. As a result, the interband transition matrix elements of the driving field are directly linked to the quantum metric through the structure of the Bloch eigenstates, culminating in the key conclusion that the Floquet band-gap renormalization is fully determined by the quantum metric.

The negative sign in Eq.~(\ref{eq:magnus_main}) implies $\mathcal{H}^{(2)}_{\sigma_z} < 0$, reducing the effective $D_s$ parameter. Consequently, the valence band maximum shifts \textit{upward} by $|\mathcal{H}^{(2)}_{\sigma_z}|$ and the conduction band minimum shifts \textit{downward} by the same amount, closing the gap symmetrically from both sides. This symmetric closure is a direct consequence of the $\sigma_z$-only structure of the Magnus correction---a feature absent in generic Floquet systems where $1/\omega$ Berry-curvature terms typically break particle-hole symmetry.

At small $\alpha$, the Bessel function $J_1(\alpha) \approx \alpha/2$ dominates the sum, giving $\Delta E_g \propto g_{xx} \alpha^2$, which establishes the quantum metric as the leading coupling channel. For larger $\alpha$ (intermediate driving), the full Bessel series in Eq.~(\ref{eq:magnus_gap}) must be retained. For $\alpha \gtrsim \pi$, where the Magnus series itself may diverge, we employ exact numerical Floquet diagonalization as the non-perturbative reference.

\begin{figure*}[htbp]
\centering
\includegraphics[width=0.95\textwidth]{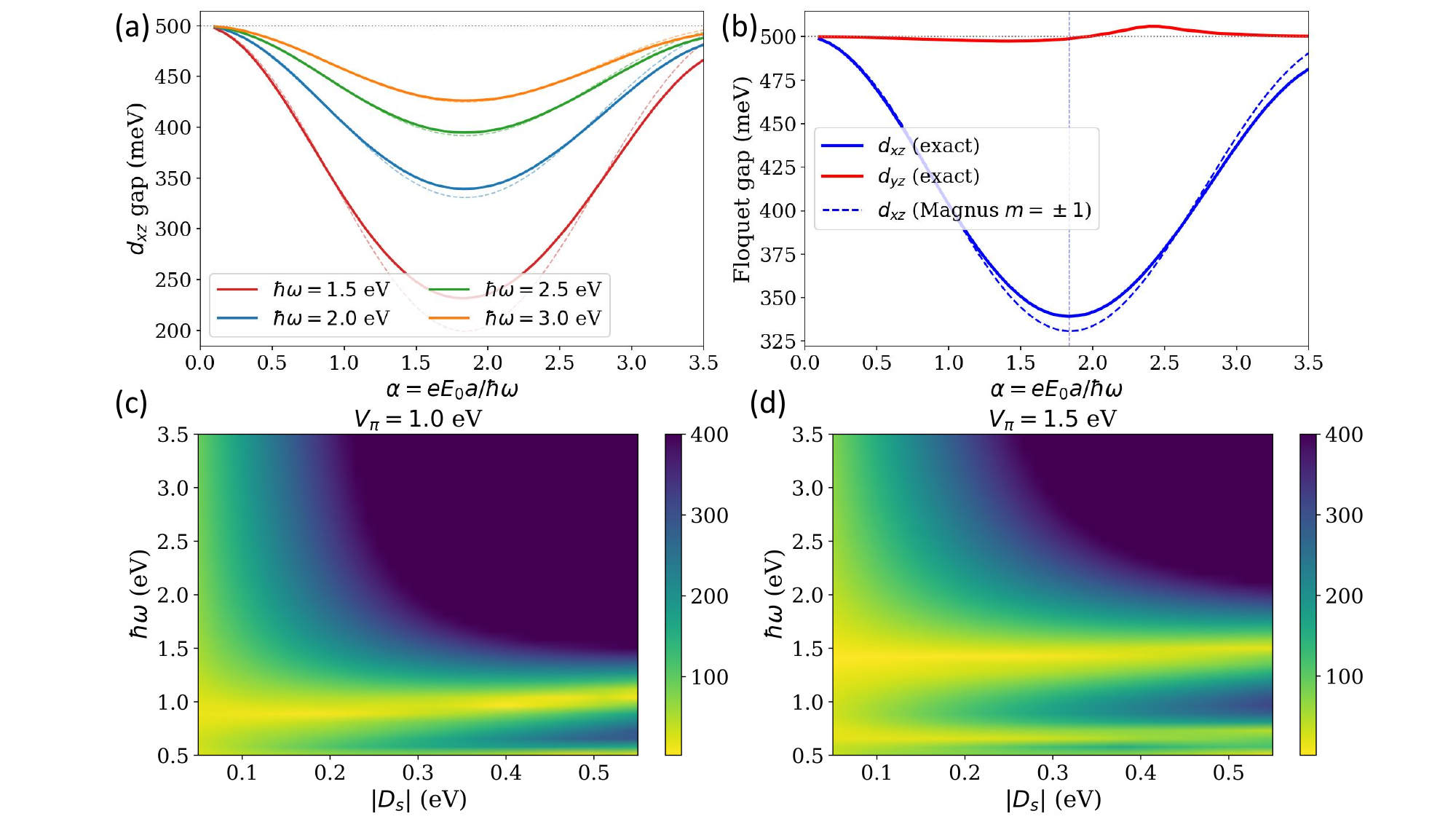}
\caption{Floquet gap evolution and Floquet phase diagram. (a) Floquet bandgaps of $d_{xz}$ orbital for different photon energies and different $\alpha$ values. Solid lines refer to the numerical results, and dashed lines represent the Magnus results. Here $V_\pi=1.0$\,eV, $V_\delta=0.15$\,eV, and $|D_s|=0.25$\,eV. (b) Comparison of Floquet bands for $d_{xz}$ orbital and $d_{yz}$ orbital. Here photon energy is $\hbar \omega=0.2$\,eV. Floquet phase diagram with $\alpha = 1.87$ in the $(|D_s|,\hbar\omega)$ plane for (c) $V_\pi=1.0$\,eV, $V_\delta=0.15$\,eV, and (d) $V_\pi=1.5$\,eV, $V_\delta=0.15$\,eV.}
\label{fig:phase_diagram}
\end{figure*}

\textit{Quantum-metric-driven light-induced ferrovalley phase.} The gap renormalization derived above produces a light-induced ferrovalley phase (illustrated schematically in Fig.~1), with antiferromagnetic order preserved throughout. At low intensity both orbitals remain gapped but their gaps evolve asymmetrically: the $d_{yz}$ channel, coupling through $V_\delta\ll V_\pi$, is barely affected, while the $d_{xz}$ gap shrinks under Floquet dressing.  This valley gap imbalance defines the ferrovalley phase: the system remains insulating yet already exhibits pronounced valley polarization whose magnitude directly encodes $g_{xx}^{xz}/g_{xx}^{yz}\propto(V_\pi/V_\delta)^2$. 

Exact Floquet diagonalization with $N_{\rm ph}=5$ photon sectors yields for representative parameters ($V_\pi=1.0$\,eV, $V_\delta=0.15$\,eV, $|D_s|=0.25$\,eV, $\hbar\omega=2.0$\,eV) a minimum $d_{xz}$ Floquet gap of $\sim0.34$\,eV ($32\%$ reduction from $E_g^0=0.50$\,eV) at $\alpha\approx1.87$, which \emph{increases} at larger $\alpha$. The $d_{yz}$ gap remains essentially unchanged, confirming the strong orbital selectivity as shown in Fig.~2(c-d). Thus, Floquet engineering drives the altermagnet from paravalley to ferrovalley states.


Fig.~3(a–b) demonstrate that Floquet engineering with linearly polarized light along the $x$ direction is highly effective for the $d_{xz}$ orbital, and the numerical calculations are consistent with the analytical results obtained from the Magnus expansion. Our parameter scan as shown in Fig.~3(c-d) confirms that the $d_{xz}$ gap can be strongly suppressed to several meV. These results establish the ferrovalley phase and its defining orbital asymmetry as robust predictions of the quantum metric mechanism. 



\textit{Discussion and outlook.} The two analytical methods employed in this work, the Magnus expansion and the van Vleck high-frequency expansion. exhibit complementary ranges of validity. The Magnus expansion provides the primary analytical framework: the interband matrix element relation (Eq.~\ref{eq:interband_metric}) connects the Peierls-induced $k$-space modulation directly to the quantum metric, and the resulting gap correction (Eqs.~\ref{eq:magnus_main}--\ref{eq:magnus_gap}) is expressed explicitly in geometric form. The expansion remains convergent for driving amplitudes $\alpha\lesssim\pi$. The van Vleck expansion (Appendix~\ref{app:vanvleck}) is equivalent to Magnus at second order in $1/\omega$ but does not resum the driving amplitude $\alpha$ non-perturbatively, making it suitable only in the strict high-frequency limit. Exact Floquet diagonalization covers all $\alpha$ and corroborates the Magnus predictions. For the representative parameters $V_\pi=1.0$\,eV, $V_\delta=0.15$\,eV, $|D_s|=0.25$\,eV, $\hbar\omega=2.0$\,eV, exact diagonalization yields a $32\%$ gap reduction at $\alpha\approx1.87$, consistent with the Magnus prediction.

The mechanism differs from conventional Floquet gap engineering in two key respects.  First, the $1/\omega$ van Vleck correction vanishes identically because $[\mathcal{H}_{-m},\mathcal{H}_m]\propto[\sigma_x,\sigma_x]=0$---a unique consequence of the real Hamiltonian that eliminates Berry curvature coupling at all orders.  Second, the orbital selectivity is structurally determined by the hopping anisotropy. The $\omega^2$ scaling of the characteristic field is generic to second-order Floquet theory; the \emph{distinguishing} signature is the orbital asymmetry, which directly encodes the quantum metric anisotropy and is falsifiable through polarization-dependent measurements.
We now return to the question raised in the Introduction: is quantum metric coupling under Floquet driving specific to d-wave altermagnets, or is it a generic consequence of vanishing Berry curvature? The answer is that such coupling is \emph{generic} whenever Berry curvature vanishes by symmetry. In any real Hamiltonian, the $1/\omega$ term (curvature-coupled) is absent and the leading $1/\omega^2$ correction couples exclusively to the metric. The d-wave altermagnets are special not because the coupling is unique, but because they uniquely combine vanishing curvature with broken spin degeneracy, making the ferrovalley phase detectable through spin-resolved probes. For instance, centrosymmetric semiconductors with negligible spin-orbit coupling (e.g., monolayer h-BN or group-IV monochalcogenides) possess real Hamiltonians and vanishing Berry curvature in broad momentum regions; under off-resonant light, their gap renormalization reduces to the same quantum metric form derived here, though spin degeneracy prevents spin-resolved ferrovalley detection. The spin-orbital locking in d-wave altermagnets, where the opposite spin polarization of the two valleys is enforced by the staggered exchange field, is what elevates the generic metric coupling into an experimentally observable ferrovalley signal.

A second distinctive signature of the quantum metric origin is the polarization response, which provides an independent and falsifiable experimental control.
The quantum metric $g_{xx}$ couples to the electric field through the $k_x$ derivative of the Hamiltonian: $\partial_{k_x}C^{xz} \propto V_\pi\sin(k_x a)$ for $d_{xz}$ and $\partial_{k_x}C^{yz} \propto V_\delta\sin(k_x a)$ for $d_{yz}$.
For $x$-polarized light, the Peierls substitution modulates $k_x$, so the Floquet gap correction for the $d_{xz}$ channel is proportional to $g_{xx}^{xz} \propto V_\pi^2$, while for $d_{yz}$ it is proportional to $g_{xx}^{yz} \propto V_\delta^2$.
The ratio of gap reductions is therefore $(V_\delta/V_\pi)^2 \sim 10^{-2}$, meaning the $d_{yz}$ gap is essentially unaffected by $x$-polarized light.

For $y$-polarized light, the roles are reversed: the Peierls substitution modulates $k_y$, coupling to $\partial_{k_y}C^{xz} \propto V_\delta\sin(k_y a)$ and $\partial_{k_y}C^{yz} \propto V_\pi\sin(k_y a)$.
Now the $d_{yz}$ gap is strongly suppressed while $d_{xz}$ remains nearly unchanged.
This 90$^\circ$ polarization rotation interchanges which valley gap is suppressed---a dramatic and unambiguous signature.
If the observed ferrovalley signal indeed originates from the quantum metric mechanism, the following must hold: (i)~the suppressed valley must switch from $d_{xz}$ to $d_{yz}$ upon rotating the pump polarization from $\hat{x}$ to $\hat{y}$, and (ii)~the magnitude of the gap reduction must scale with the square of the corresponding hopping integral ($V_\pi^2$ or $V_\delta^2$).



Beyond the polarization anisotropy, the ferrovalley phase proposed here is directly accessible to several established experimental probes. Femtosecond time-resolved spin-resolved ARPES (tr-ARPES) can directly image the valley-resolved band structure with sub-eV energy resolution and femtosecond temporal resolution, enabling the measurement of both the gap reduction and the valley gap difference $\Delta E_g^{xz-yz}$ as functions of pump fluence and polarization. The orbital selectivity---$d_{xz}$ gap strongly suppressed while $d_{yz}$ remains nearly unchanged---provides a distinctive fingerprint: the ARPES spectrum should exhibit a pronounced splitting between the two valley gaps that grows with increasing pump field. Optical pump-probe spectroscopy offers a complementary probe: the differential reflectivity or transmission $\Delta R/R$ at photon energies near the $d_{xz}$ gap edge is expected to show a red-shift proportional to the quantum metric $g_{xx}$, while the $d_{yz}$ gap edge remains stationary, yielding a differential signal that vanishes under $y$-polarized pumping---a null test that unambiguously isolates the quantum metric origin. 


In conclusion, we have demonstrated that off-resonant light drives a purely quantum metric-mediated band-gap renormalization in d-wave altermagnets, realizing a ferrovalley phase whose valley gap difference directly encodes the quantum metric anisotropy with no Berry curvature contribution at any order. The orbital asymmetry, together with its dramatic polarization dependence---where a $90^\circ$ rotation of the pump polarization interchanges which valley gap is suppressed---provides a distinctive and falsifiable signature of pure quantum geometry, accessible to femtosecond spin-resolved ARPES and optical pump-probe spectroscopy. Beyond establishing d-wave altermagnets as a pristine platform where quantum metric effects can be isolated, controlled by light, and read out through valley polarization, this work opens a new direction for quantum geometry in Floquet-engineered materials: whenever Berry curvature vanishes by symmetry, the leading light-matter coupling reduces to a universal quantum-metric form, with the material-specific spin-orbital structure determining the experimental observability of the resulting geometric phases.

\section{Acknowledgments}
This work was supported by the National Natural Science Foundation of China (No. 12304217), the National Key Research and Development Program of China (No. 2024YFA1410300), the Natural Science Foundation of Hunan Province (No. 2025JJ60002), and the Fundamental Research Funds for the Central Universities from China (No. 531119200247).

\clearpage
\appendix
\onecolumngrid
\section{Supplementary materials for "Quantum-metric-driven light-induced ferrovalley state in d-wave altermagnets"}\label{app:details}
\setcounter{equation}{0}
\renewcommand{\theequation}{S\arabic{equation}}
\subsection{A.~Quantum metric calculation}\label{app:qmetric}

For a real two-band Hamiltonian $\mathcal{H}=h_x\sigma_x+h_z\sigma_z$ with $h=\sqrt{h_x^2+h_z^2}$ and $\tan\theta=h_x/h_z$, the normalized eigenstates are $|u_+\rangle=(\cos\frac{\theta}{2},\sin\frac{\theta}{2})^T$, $|u_-\rangle=(-\sin\frac{\theta}{2},\cos\frac{\theta}{2})^T$. The quantum metric is
\begin{equation}
g_{\mu
u}(k) = {\rm Re}\left[\langle\partial_{k_\mu}u_-|u_+\rangle\langle u_+|\partial_{k_
u}u_-\rangle\right].
\end{equation}
Computing $\langle u_+|\partial_{k_x}u_-\rangle = -\frac{1}{2}\partial_{k_x}\theta$ and $\partial_{k_x}\theta = (h_z/h^2)\partial_{k_x}h_x$ gives
\begin{equation}
g_{xx}(k) = \frac{D_s^2 [\partial_{k_x}C(k)]^2}{4h^4(k)}.
\end{equation}
For the $d_{xz}$ orbital, $\partial_{k_x}C^{xz} = -a V_\pi\sin(k_x a)$, so $g_{xx}^{xz}=a^2 D_s^2 V_\pi^2\sin^2(k_x a)/(4h^4)$. For $d_{yz}$, $g_{xx}^{yz}=a^2 D_s^2 V_\delta^2\sin^2(k_x a)/(4h^4)$. Since $V_\pi\gg V_\delta$, $g_{xx}^{xz}\gg g_{xx}^{yz}$. The Berry curvature vanishes: $\Omega_{xy}=2{\rm Im}[\langle\partial_{k_x}u_-|u_+\rangle\langle u_+|\partial_{k_y}u_-\rangle]=0$ because all matrix elements are real for a real Hamiltonian.

The interband matrix element of the Hamiltonian gradient follows directly from the eigenstates.
Using $\langle u_-|\sigma_x|u_+\rangle = \cos\theta = h_z/h$ and $\langle u_-|\sigma_z|u_+\rangle = -\sin\theta = -h_x/h$, together with $\partial_{k_\mu}\mathcal{H} = (\partial_{k_\mu}h_x)\sigma_x + (\partial_{k_\mu}h_z)\sigma_z$, we find
\begin{equation}
\langle u_-|\partial_{k_\mu}\mathcal{H}|u_+\rangle = \frac{h_z\partial_{k_\mu}h_x - h_x\partial_{k_\mu}h_z}{h}.
\label{eq:dH_matrix}
\end{equation}
Recognizing $\partial_{k_\mu}\theta = (h_z\partial_{k_\mu}h_x - h_x\partial_{k_\mu}h_z)/h^2$, this simplifies to $\langle u_-|\partial_{k_\mu}\mathcal{H}|u_+\rangle = h\,\partial_{k_\mu}\theta$.
Since $g_{\mu\mu} = (\partial_{k_\mu}\theta)^2/4$, we obtain
\begin{equation}
|\langle u_-|\partial_{k_\mu}\mathcal{H}|u_+\rangle|^2 = 4h^2(k)\,g_{\mu\mu}(k),
\end{equation}
which is the relation used in the main text [Eq.~(\ref{eq:interband_metric})] to connect the Magnus Fourier components to the quantum metric.

\subsection{B.~Magnus expansion}\label{app:magnus}

The Magnus expansion \cite{Magnus1954,Blanes2009,Martinez2003} constructs the effective static Floquet Hamiltonian $\mathcal{H}_F$ from the time-ordered exponential:
\begin{equation}
U(T,0) = \mathcal{T}\exp\!\left[-\frac{i}{\hbar}\int_0^T \mathcal{H}(t)\,dt\right] \equiv \exp\!\left[-\frac{i}{\hbar}\mathcal{H}_F T\right].
\end{equation}
Physically, $\mathcal{H}_F$ is the \emph{exact} time-independent Hamiltonian whose evolution over one period reproduces the full time-dependent dynamics at stroboscopic times. The Magnus expansion expresses $\mathcal{H}_F$ as a series $\mathcal{H}_F = \mathcal{H}^{(0)} + \mathcal{H}^{(1)} + \mathcal{H}^{(2)} + \cdots$ where $\mathcal{H}^{(n)}$ involves $n$ nested commutators and $n+1$ time integrals. The Magnus expansion resums the Bessel-function dependence $J_n(\alpha)$ non-perturbatively and has a convergence radius $\alpha \lesssim \pi$, substantially larger than the $\alpha \ll 1$ range accessible to linear expansions.

The first two Magnus terms are:
\begin{align}
\mathcal{H}^{(0)} &= \frac{1}{T}\int_0^T dt\,\mathcal{H}(t), \label{eq:magnus0}\\
\mathcal{H}^{(1)} &= \frac{1}{2iT}\int_0^T dt_1\int_0^{t_1} dt_2\,[\mathcal{H}(t_1), \mathcal{H}(t_2)]. \label{eq:magnus1}
\end{align}

For our Hamiltonian $\mathcal{H}(k,t) = h_x\sigma_x + D_s\sigma_z$ with $h_x(t) = V_\pi\cos(k_x a + \alpha\sin\omega t) + V_\delta\cos k_y a$ and $\alpha \equiv eE_0 a/\hbar\omega$, the Jacobi-Anger expansion $\cos(A + \alpha\sin\omega t) = \sum_n J_n(\alpha)\cos(A + n\omega t)$ yields:
\begin{equation}
\mathcal{H}^{(0)}(k) = \bigl[V_\pi J_0(\alpha)\cos k_x a + V_\delta\cos k_y a\bigr]\sigma_x + D_s\sigma_z.
\end{equation}
This describes \emph{dynamical band narrowing}: the effective $\pi$-hopping is reduced by $J_0(\alpha)$. At the valence band maximum or conduction band minimum, $h_x^{\rm eff}=0$ and the gap is $2|D_s|$ for all $\alpha$. Zeroth-order Magnus never changes the gap.

The first-order commutator gives $[\mathcal{H}(t_1),\mathcal{H}(t_2)] = -2iD_s[h_x(t_1)-h_x(t_2)]\sigma_y$, producing an effective $\sigma_y$ term $\propto D_s\alpha V_\pi\sin k_x a/(\hbar\omega)$ that vanishes at the valence band maximum or conduction band minimum. First-order Magnus does not affect the gap either.

The second-order Magnus Hamiltonian is:
\begin{equation}
\mathcal{H}^{(2)} = -\frac{1}{6T}\int_0^T dt_1\int_0^{t_1} dt_2\int_0^{t_2} dt_3\,
\bigl([\mathcal{H}(t_1),[\mathcal{H}(t_2),\mathcal{H}(t_3)]] + [\mathcal{H}(t_3),[\mathcal{H}(t_2),\mathcal{H}(t_1)]]\bigr).
\label{eq:magnus2}
\end{equation}

Evaluating the nested commutators for $\mathcal{H}(t) = h_x(t)\sigma_x + D_s\sigma_z$:
\begin{align}
[\mathcal{H}(t_2),\mathcal{H}(t_3)] &= -2iD_s\bigl[h_x(t_2)-h_x(t_3)\bigr]\sigma_y, \\
[\mathcal{H}(t_1),[\mathcal{H}(t_2),\mathcal{H}(t_3)]] &= 4D_s\bigl[h_x(t_2)-h_x(t_3)\bigr]\bigl(h_x(t_1)\sigma_z - D_s\sigma_x\bigr), \\
[\mathcal{H}(t_3),[\mathcal{H}(t_2),\mathcal{H}(t_1)]] &= 4D_s\bigl[h_x(t_2)-h_x(t_1)\bigr]\bigl(h_x(t_3)\sigma_z - D_s\sigma_x\bigr).
\end{align}
Summing both contributions and retaining the $\sigma_z$ part (which governs gap renormalization):
\begin{equation}
[\mathcal{H}(t_1),[\mathcal{H}(t_2),\mathcal{H}(t_3)]] + [\mathcal{H}(t_3),[\mathcal{H}(t_2),\mathcal{H}(t_1)]]
= 4D_s\bigl[h_x(t_2)(h_x(t_1)+h_x(t_3)) - 2h_x(t_1)h_x(t_3)\bigr]\sigma_z + \mathcal{O}(\sigma_x).
\end{equation}

In the Fourier representation $\mathcal{H}(t) = \sum_n \mathcal{H}_n e^{in\omega t}$, for systems where all Fourier components $\mathcal{H}_m$ ($m 
\neq 0$) commute with each other, as in our case where $\mathcal{H}_m\propto\sigma_x$, the Magnus $\mathcal{H}^{(2)}$ takes the form:
\begin{equation}
\mathcal{H}^{(2)} = \sum_{m
\neq 0} \frac{[[\mathcal{H}_{-m}, \mathcal{H}_0], \mathcal{H}_m]}{(m\hbar\omega)^2}
+ \frac{1}{2}\sum_{m
\neq 0} \frac{[\mathcal{H}_{-m}, [\mathcal{H}_{0}, \mathcal{H}_{m}]]}{(-m^2)(\hbar\omega)^2}.
\end{equation}
The first sum is the standard van Vleck $1/\omega^2$ term \cite{Eckardt2015}. The second sum is an additional Magnus contribution from the $m+n=0$ ($n=-m$) channels in the triple time integral, which is \emph{not} captured by the van Vleck expansion. Evaluating:
\begin{align}
[\mathcal{H}_{-m}, \mathcal{H}_0] &= D_s h_{-m}[\sigma_x,\sigma_z] = -2i D_s h_{-m}\sigma_y, \\
[[\mathcal{H}_{-m}, \mathcal{H}_0], \mathcal{H}_m] &= [-2i D_s h_{-m}\sigma_y,\; h_m\sigma_x] 
= -4 D_s |h_m|^2\sigma_z.
\end{align}
The second sum gives $[\mathcal{H}_{-m}, [\mathcal{H}_{0}, \mathcal{H}_{m}]] = -4 D_s |h_m|^2\sigma_z$ with prefactor $-1/(2m^2\omega^2)$, contributing $+2 D_s |h_m|^2/(m\hbar\omega)^2\sigma_z$. Summing both contributions:
\begin{equation}
`\mathcal{H}^{(2)}_{\sigma_z} = -\sum_{m\neq 0} \frac{2 D_s |\mathcal{H}_m|^2}{(m\hbar\omega)^2}\,\sigma_z.
\end{equation}

We now show that the gap correction $\mathcal{H}^{(2)}_{\sigma_z}$ is naturally expressed through the quantum metric. The key steps are: (i)~the Magnus double commutator selects the interband matrix element $|\langle u_-|\mathcal{H}_m|u_+\rangle|^2$; (ii)~at the valence band maximum or conduction band minimum, $\mathcal{H}_m$ is exactly proportional to $\partial_{k_x}\mathcal{H}$; (iii)~the interband matrix element of $\partial_{k_x}\mathcal{H}$ equals $4h^2 g_{xx}$.

In the band eigenbasis $|u_\pm\rangle$, the static Hamiltonian is diagonal: $\mathcal{H}_0 = \mathrm{diag}(h,-h)$.
The Fourier components $\mathcal{H}_m \propto \sigma_x$ have both diagonal and off-diagonal parts in this basis, with matrix elements
\begin{align}
\langle u_+|\mathcal{H}_m|u_+\rangle &= h_m\,\frac{h_x}{h}, &
\langle u_-|\mathcal{H}_m|u_-\rangle &= -h_m\,\frac{h_x}{h}, \\
\langle u_-|\mathcal{H}_m|u_+\rangle &= h_m\,\frac{D_s}{h}, &
\langle u_+|\mathcal{H}_m|u_-\rangle &= h_m\,\frac{D_s}{h},
\end{align}
where $h_m$ is the scalar coefficient of $\sigma_x$ in $\mathcal{H}_m$.
At the valence band maximum or conduction band minimum, the diagonal elements vanish and $\mathcal{H}_m$ is purely off-diagonal: $|\langle u_-|\mathcal{H}_m|u_+\rangle| = |h_m|$.
The double commutator governing $\mathcal{H}^{(2)}_{\sigma_z}$ evaluates to (see the Magnus derivation above):
\begin{equation}
[[\mathcal{H}_{-m}, \mathcal{H}_0], \mathcal{H}_m] \to -4D_s\,|\langle u_-|\mathcal{H}_m|u_+\rangle|^2\,\sigma_z,
\end{equation}
confirming that the \emph{interband} matrix element---not the bare coefficient $|\mathcal{H}_m|^2$---is the physically relevant quantity.
The diagonal matrix elements $\langle u_\pm|\mathcal{H}_m|u_\pm\rangle$ do not contribute to the $\sigma_z$ gap correction; they generate intraband energy shifts that cancel between $\pm m$ channels.

For odd $m$, the Fourier component is $\mathcal{H}_m^{\rm odd}=i\,\mathrm{sgn}(m)V_\pi J_{|m|}(\alpha)\sin(k_x a)\,\sigma_x$.
Meanwhile, $\partial_{k_x}\mathcal{H} = -a V_\pi\sin(k_x a)\,\sigma_x$ (since $\partial_{k_x}D_s=0$).
Comparing coefficients gives the exact relation, valid at all $k$:
\begin{equation}
\mathcal{H}_m^{\rm odd} = -i\,\mathrm{sgn}(m)\,\frac{J_{|m|}(\alpha)}{a}\;\partial_{k_x}\mathcal{H}.
\label{eq:Hm_to_dH}
\end{equation}

The interband matrix element of $\partial_{k_x}\mathcal{H}$ is evaluated in Appendix~\ref{app:qmetric} [Eq.~(\ref{eq:dH_matrix})], yielding
\begin{equation}
|\langle u_-|\partial_{k_x}\mathcal{H}|u_+\rangle|^2 = 4h^2(k)\,g_{xx}(k).
\end{equation}
Combining with Eq.~(\ref{eq:Hm_to_dH}):
\begin{equation}
|\langle u_-|\mathcal{H}_m|u_+\rangle|^2 = \frac{J_m^2(\alpha)}{a^2}\;4h^2(k)\,g_{xx}(k).
\label{eq:interband_to_metric}
\end{equation}

Substituting Eq.~(\ref{eq:interband_to_metric}) into $\mathcal{H}^{(2)}_{\sigma_z}$ [Eq.~(\ref{eq:magnus_main})] and projecting onto the gap at the valence band maximum or conduction band minimum ($h=|D_s|$):
\begin{equation}
\delta D_s = -8D_s^3\,g_{xx}\sum_{m\neq 0}\frac{J_m^2(\alpha)}{a^2(m\hbar\omega)^2},
\end{equation}
which is Eq.~(\ref{eq:magnus_gap}) of the main text.
The gap renormalization is thus expressed exclusively through the quantum metric. 

In summary, the Peierls substitution enacts a periodic translation of the electronic crystal momentum in $k$-space, for which the $k$-space gradient operator $\partial_k$ serves as the infinitesimal generator. It follows rigorously that the Fourier components of the time-periodic Hamiltonian are proportional to the $k$-space derivatives of the static Hamiltonian. The interband transition matrix elements of the driving field are then directly tied to the quantum metric through the structure of the Bloch eigenstates, leading to the central result that the Floquet band-gap renormalization is entirely governed by the quantum metric. The general derivation of quantum-metric-driven band-gap renormalization is presented in Section~D.

At the valence band maximum or conduction band minimum and small $\alpha$, $J_1(\alpha) \approx \alpha/2$, and the Magnus result (including both $\pm m$ channels) reduces to $\delta D_s^{(2)} \approx -D_s(eE_0 a V_\pi)^2/(\hbar\omega)^4$, establishing the $\alpha^2$ scaling characteristic of second-order Floquet physics. For representative parameters ($V_\pi=1.0$\,eV, $D_s=0.25$\,eV, $\hbar\omega=2.0$\,eV, $\alpha=2.0$), the second-order Magnus estimate gives $D_s^{\rm eff} \approx 0.167$\,eV, i.e., $E_g^{\rm Magnus\,(2)} \approx 0.334$\,eV ($67\%$ of the zero-field gap), in excellent agreement with the exact Floquet minimum of $0.339$\,eV. Residual differences arise from higher Magnus orders $H^{(3)}, H^{(4)},\dots$ representing multi-photon virtual transitions.


\subsection{C.~van Vleck high-frequency expansion}\label{app:vanvleck}

The van Vleck expansion \cite{Eckardt2015,Bukov2015} is a systematic $1/\omega$ expansion of the effective Floquet Hamiltonian constructed via a unitary transformation that block-diagonalizes the Floquet matrix:
\begin{equation}
\mathcal{H}_{\rm eff} = \mathcal{H}_0 + \sum_{m
\neq 0}\frac{[\mathcal{H}_{-m},\mathcal{H}_m]}{2m\hbar\omega}
+ \sum_{m
\neq 0}\frac{[[\mathcal{H}_{-m},\mathcal{H}_0 - m\hbar\omega],\mathcal{H}_m]}{2(m\hbar\omega)^2}
+ \mathcal{O}(1/\omega^3). \label{eq:vanvleck}
\end{equation}

It is important to clarify the distinctions among the three analytical methods. The Magnus expansion resums the Bessel-function dependence $J_n(\alpha)$ non-perturbatively through the Jacobi-Anger identity. Valid for $\alpha \lesssim \pi$, it captures both the leading $\alpha^2$ gap renormalization and higher-order multi-photon contributions at the characteristic field strengths ($\alpha \sim 1$--$2$) relevant for experiments.

Van Vleck expansion is a $1/\omega$ high-frequency expansion constructing an effective static Hamiltonian order by order. At second order in $1/\omega$, it yields the same $\alpha^2$ functional dependence as the Magnus result when both are expanded to the same order in $\alpha$. However, the van Vleck $1/\omega$ term ($\propto \sum_m [\mathcal{H}_{-m},\mathcal{H}_m]/(m\hbar\omega)$) couples to the Berry curvature in generic systems and vanishes identically for our real Hamiltonian---a distinctive advantage of d-wave altermagnets.

Magnus expansion constructs the effective Floquet Hamiltonian directly from the time-ordered exponential without expanding in either $\alpha$ or $1/\omega$. It has the largest convergence radius ($\alpha \lesssim \pi$) and resums the Bessel-function dependence $J_n(\alpha)$ non-perturbatively. At second order, Magnus $\mathcal{H}^{(2)}$ contains the van Vleck $1/\omega^2$ term plus an additional triple-integral contribution absent from the van Vleck expansion.

The Magnus expansion is applicable in a wider range of situations than the van Vleck expansion. In the main text, results are presented first within the Magnus framework, then refined by exact Floquet diagonalization, with the perturbative small-$\alpha$ limit discussed as a consistency check.

The Jacobi-Anger Fourier components are:
\begin{align}
\mathcal{H}_{n=0} &= [V_\pi J_0(\alpha)\cos k_x a + V_\delta\cos k_y a]\sigma_x + D_s\sigma_z, \\
\mathcal{H}_{n\;{\rm even},
\neq 0} &= V_\pi J_{|n|}(\alpha)\cos k_x a\,\sigma_x, \\
\mathcal{H}_{n\;{\rm odd}} &= i\,\mathrm{sgn}(n)\,V_\pi J_{|n|}(\alpha)\sin k_x a\,\sigma_x.
\end{align}

Since all $\mathcal{H}_m \propto \sigma_x$ (for $m 
\neq 0$), the $1/\omega$ commutator $[\mathcal{H}_{-m},\mathcal{H}_m] \propto [\sigma_x,\sigma_x] = 0$ vanishes identically. This is a distinctive feature of d-wave altermagnets: \emph{the leading high-frequency correction couples exclusively to the quantum metric, with vanishing Berry curvature contribution.} At small $\alpha$, the $1/\omega^2$ double-commutator has the same $\alpha^2$ functional scaling as the Magnus result.

\section{D.~General Derivation of Quantum-Metric-Driven Band-Gap Renormalization}

We start from a generic single-particle Bloch Hamiltonian $\hat{H}_0(\boldsymbol{k})$ for a periodic crystal, whose eigenstates $|u_n(\boldsymbol{k})\rangle$ and eigenenergies $E_n(\boldsymbol{k})$ satisfy
\begin{equation}
\hat{H}_0(\boldsymbol{k}) \, |u_n(\boldsymbol{k})\rangle = E_n(\boldsymbol{k}) \, |u_n(\boldsymbol{k})\rangle ,
\end{equation}
where $|u_n(\boldsymbol{k})\rangle$ denotes the periodic part of the Bloch wave function.

Under linearly polarized light along the $x$-direction, the minimal electromagnetic coupling to lattice electrons is described by the Peierls substitution in the Coulomb gauge:
\begin{equation}
k_x \to k_x + \delta k(t), \quad \delta k(t) = \frac{e A(t)}{\hbar} = k_A \sin\omega t ,
\end{equation}
where $k_A = e A_0 / \hbar$ is the amplitude of the momentum modulation, $A(t) = A_0 \sin\omega t$ is the vector potential, and $\omega$ is the driving frequency. The time-dependent driven Hamiltonian is therefore equivalent to a periodic translation of the static Hamiltonian in momentum space:
\begin{equation}
\hat{H}(\boldsymbol{k}, t) = \hat{H}_0\!\left( k_x + k_A \sin\omega t,\, k_y \right) .
\end{equation}
This relation holds rigorously for any Bloch Hamiltonian, independent of the specific lattice, orbital, or hopping structure.

Projected onto the eigenbasis of $\hat{H}_0$, second-order Magnus Hamiltonian reduces to the standard second-order perturbation formula describing virtual interband transitions. For a given band $n$, the second-order energy shift reads:
\begin{equation}
\Delta E_n^{(2)} = \sum_{m \neq 0} \sum_{l \neq n} \frac{ \left| \langle u_n | \hat{H}_m | u_l \rangle \right|^2 }{ E_n - E_l + m \hbar \omega } .
\end{equation}

For an arbitrary Hamiltonian, we expand it in a Taylor series around $k_x$, which is valid for weak driving $k_A \ll 1/a$:
\begin{equation}
\hat{H}_0(k_x + k_A \sin\omega t) = \hat{H}_0(k_x) + k_A \sin\omega t \cdot \partial_{k_x} \hat{H}_0 + \mathcal{O}(k_A^2) .
\end{equation}
Then we directly extract the first-order Fourier components ($m=\pm1$):
\begin{equation}
\hat{H}_{\pm 1}(\boldsymbol{k}) = \pm \frac{k_A}{2i} \, \partial_{k_x} \hat{H}_0(\boldsymbol{k}) .
\end{equation}

This is a universally exact result at leading order in the driving amplitude: the first Fourier component of a Peierls-driven Hamiltonian is strictly proportional to the k-space gradient of the static Hamiltonian, with a proportionality constant set solely by the driving amplitude. Physically, this reflects the fact that $\partial_{k_x}$ is the infinitesimal generator of translations in momentum space.

We now relate the interband matrix element of the Hamiltonian gradient $\langle u_l | \partial_{k_x} \hat{H}_0 | u_n \rangle$ to the geometric structure of the Bloch states. Differentiating the eigenvalue equation (1) with respect to $k_x$:
\begin{equation}
(\partial_{k_x} \hat{H}_0) |u_n\rangle + \hat{H}_0 \, \partial_{k_x} |u_n\rangle = (\partial_{k_x} E_n) |u_n\rangle + E_n \, \partial_{k_x} |u_n\rangle .
\end{equation}
Left-multiplying by $\langle u_l |$ with $l \neq n$, and using orthonormality $\langle u_l | u_n \rangle = 0$ together with $\langle u_l | \hat{H}_0 = E_l \langle u_l |$, we obtain the fundamental Bloch identity:
\begin{equation}
\langle u_l | \partial_{k_x} \hat{H}_0 | u_n \rangle = (E_n - E_l) \cdot \langle u_l | \partial_{k_x} u_n \rangle .
\end{equation}
This relation holds for any band structure and connects the interband matrix element of the Hamiltonian gradient directly to the overlap between gradients of Bloch eigenstates.

A core premise of this work is that the Hamiltonian is entirely real, so the Bloch eigenstates can be chosen to be real-valued everywhere in the Brillouin zone. In this case, $\langle u_l | \partial_{k_\mu} u_n \rangle$ is purely real, the Berry curvature vanishes identically ($\Omega_{\mu\nu}^{(n)} \equiv 0$), and the quantum metric reduces to
\begin{equation}
g_{xx}^{(n)} = \sum_{l \neq n} \left| \langle u_l | \partial_{k_x} u_n \rangle \right|^2 .
\end{equation}

For a two-band system (valence band $v$, conduction band $c$) with zero-field gap $E_g^0 = E_c - E_v$, the sum contains only one term. Combining Eqs.~(10) and (12) gives the central geometric identity:
\begin{equation}
\left| \langle u_c | \partial_{k_x} \hat{H}_0 | u_v \rangle \right|^2 = (E_g^0)^2 \cdot g_{xx}^{(v)} .
\end{equation}

We now evaluate the gap correction at the valence band maximum or conduction band minimum, where $\partial_{k_x} E = 0$ so the first-order intraband energy shift vanishes. In the far-off-resonant limit $\hbar\omega \gg E_g^0$, only $m=\pm1$ contribute significantly. The second-order energy shift is
\begin{align}
\Delta E_v^{(2)} &= \frac{ \left| \langle u_v | \hat{H}_1 | u_c \rangle \right|^2 }{ E_v - E_c + \hbar\omega } + \frac{ \left| \langle u_v | \hat{H}_{-1} | u_c \rangle \right|^2 }{ E_v - E_c - \hbar\omega } \nonumber \\
&= \frac{k_A^2}{4} \left| \langle u_c | \partial_{k_x} \hat{H}_0 | u_v \rangle \right|^2 \cdot \left( \frac{1}{-E_g^0 + \hbar\omega} + \frac{1}{-E_g^0 - \hbar\omega} \right) \nonumber \\
&\approx -\frac{k_A^2 \, (E_g^0)^3}{2 \, (\hbar\omega)^2} \, g_{xx}^{(v)} .
\end{align}

Owing to the particle-hole symmetric structure of the real two-band Hamiltonian, the conduction band shifts downward by the same amount. The total band-gap renormalization is therefore
\begin{equation}
\Delta E_g = \Delta E_c - \Delta E_v = - \frac{ k_A^2 \, (E_g^0)^3 }{ (\hbar\omega)^2 } \, g_{xx}^{(v)} .
\end{equation}

Thus, the second-order Floquet band-gap correction is directly and exclusively proportional to the quantum metric, with zero Berry curvature contribution at all orders. This derivation relies only on the Peierls substitution, second-order Magnus perturbation theory, and the reality of the Hamiltonian; it is independent of the specific lattice, orbital, or hopping form. The d-wave altermagnet tight-binding model studied in the main text is a concrete material realization of this universal quantum-geometric mechanism.

\end{document}